\documentstyle[12pt]{article}
\def\be {\begin{equation}}
\def\ee {\end{equation}}
\def\ba {\begin{eqnarray}}
\def\ea {\end{eqnarray}}

\def\bi {\begin{itemize}}
\def\ei {\end{itemize}}
\begin{document}
\def\bea{\begin{eqnarray}}
\def\eea{\end{eqnarray}}
\title{\bf {Space Noncommutativity Corrections to the Cardy-Verlinde Formula }}
 \author{M.R. Setare  \footnote{E-mail: rezakord@ipm.ir}
  \\{Physics Dept. Inst. for Studies in Theo. Physics and
Mathematics(IPM)}\\
{P. O. Box 19395-5531, Tehran, IRAN }}
\date{\small{}}

\maketitle
\begin{abstract}
In this letter we compute the corrections to the Cardy-Verlinde
formula of  Schwarzschild  black holes. These corrections stem
from the space noncommutativity. Because the Schwarzschild black
holes are non rotating, to the first order of perturbative
calculations, there is no any effect on the properties of black
hole due to the noncommutativity of space.
 \end{abstract}
\newpage

 \section{Introduction}
The Cardy-Verlinde formula proposed
   by Verlinde \cite{Verl}, relates the entropy of a  certain CFT with its total
energy and its Casimir energy in arbitrary dimensions. Using the
AdS$_{d}$/CFT$_{d-1}$ \cite{AdS} and dS$_{d}$/CFT$_{d-1}$
correspondences \cite{AS} , this formula has been shown to hold
exactly for different black
holes (see for example {\cite{odi}-\cite{set2}}).\\
Black hole thermodynamic quantities depend on the Hawking
temperature via the usual thermodynamic relations. The Hawking
temperature undergoes corrections from many sources:the quantum
corrections \cite{das}, the self-gravitational corrections
\cite{kk}, and the
corrections due to the generalized uncertainty principle\cite{set6}.\\
In this letter we concentrate on the corrections due to the space
non commutativity. Recently there has been considerable interest
in the possible effects of the non commutative space \cite{ws}.
In \cite{ah} the author have argued that every consideration on
space time measurement that allows gravitational effects asks for
non-commutative space time.
\\
By considering a black hole as quantum state instead of a
classical object \cite{toh} and according to quantum mechanics
principle, one can conclude its energy and its corresponding
conjugate time can not be simultaneously measured exactly. The
energy of states should approximately be the hole's gravitational
energies measured at the region of the horizon. In other hand the
gravitational energies are quasilocal, in the Schwarzschild black
hole case, this quasilocal energy is proportional to the radius
of event horizon. Therefore the uncertainty relation between
energy and time in the event horizon region lead that the radial
coordinate is noncommutative with time at the horizon.\\
In section $2$ we drive the corrections to the thermodynamic
quantities due to the space noncommutativity. In section $3$ we
consider the generalized Cardy-Verlinde formula of a
$4-$dimensional Schwarzschild black hole\cite{klem,yum}, then we
obtain the space noncommutativity corrections to this entropy
formula.

\section{Schwarzschild black hole in noncommutative space}
A $4-$dimensional Schwarzschild black hole of mass $M$ is
described by the metric
\begin{equation}\label{metr}
ds^{2}=-(1-\frac{2M}{r})dt^2+(1-\frac{2M}{r})^{-1}dr^2+r^2(d\theta^{2}+\sin^{2}\theta
d\varphi^{2})
\end{equation}
Considering the black hole as quantum state, the energy of
quantum state is the quasi-local energy at the horizon \cite{yb}.
According to the definition given by Brown and York \cite{by} the
quasi-local energy is as
\begin{equation}\label{qe}
E_{QL}=\frac{1}{8\pi G}\oint_{\Sigma}d^2x \sqrt{\sigma}(K-K_0)
\end{equation}
where $\Sigma$ is the two dimension spherical surface, $\sigma$
is the determinant of the $2-$metric on $\Sigma$, $K$ is the
trace of the extrinsic curvature of $\Sigma$, and $K_0$ is a
reference term that is used to normalize the energy with respect
to a reference spacetime. In the Schwarzschild metric case we have
\begin{equation}\label{ext}
K_{\theta}^{\theta}=K_{\varphi}^{\varphi}=-\frac{r-2M}{r^2}
\end{equation}
By considering $K_0$ as following
\begin{equation}\label{extr}
K_0=\frac{-1}{r}
\end{equation}
We obtain
\begin{equation}\label{qein}
E_{QL}(r\rightarrow\infty)=\frac{r}{G}(1-1+\frac{2M}{r})=\frac{2M}{G}
\end{equation}
also as the horizon  $r=r_H=2M$ we have
\begin{equation}\label{qeinh}
E(r_H)=\frac{r_H}{G}
\end{equation}
 The quantum state
energy $E$ and its conjugate time $t$ can not be simultaneously
measured exactly \cite{toh, yb}. By considering $E$ and $t$ as
operators, we have
\begin{equation}\label{alg}
[t,E]=i.
\end{equation}
Using Eq.(\ref{qeinh}) we can write
\begin{equation}\label{alg1}
[t,r]|r=r_{H}=il_{p}^{2}.
\end{equation}
where $l_p=\sqrt{G}$. Due to quantum measurement effects, $r_H$
spread into a range of $(r_H-\Delta r,r_H+\Delta r )$. Therefore
we can write
\begin{equation}\label{alg2}
[t,r]|_{r\in (r_H-\Delta r,r_H+\Delta r )}=
il_{p}^{2},
\end{equation}
where $\Delta r$ is a distance about Planck length scale,
Eq.(\ref{alg2}) is the spacetime noncommutative relation. We
rewrite Eq.(\ref{alg2}) as following
\begin{eqnarray}\label{alge}
[x_{i},x_j]=i\theta_{ij}, \hspace{.3cm}(i,j=0,1), (x_0=t,x_1=x),
\hspace{.5cm} [x_k,x_{\mu}]=0 \hspace{.3cm}(k=2,3;\mu=0,1,2,3)
\end{eqnarray}
For noncommutative black hole, one can use the usual definition
for the metric
\begin{equation}\label{metrnon}
ds^{2}=-f(\tilde{r})dt^2+f(\tilde{r})^{-1}dr^2+r^2(d\theta^{2}+\sin^{2}\theta
d\varphi^{2})
\end{equation}
where $\tilde{r}$ satisfies following Poisson brackets
\begin{equation}\label{pois}
\{\tilde{x_i},\tilde{x_j}\}=\theta_{ij},
\hspace{.5cm}\{\tilde{x_i},\tilde{p_j}\}=\delta_{ij},\hspace{.5cm}\{\tilde{p_i},\tilde{p_j}\}=0.
\end{equation}
However, one should be aware that there is no modified Einstein
equation in this case\cite{li}. Since the noncommutativity
parameter should very small compared to the length scales of the
black hole, one can treat the noncommutative effects as some
perturbations of the commutative counter-part, $f(\tilde{r})$ in
terms of the noncommutative coordinates $\tilde{x_i}$ is as
\begin{equation}\label{feq}
f(\tilde{r})=1-\frac{2M}{\sqrt{\tilde{x_i}\tilde{x_i}}}
\end{equation}
We note that there is a new coordinate system \cite{chai}
\begin{equation}\label{newco}
x_i=\tilde{x_i}+\frac{1}{2}\theta_{ij}\tilde{p_j},
\hspace{1cm}p_j=\tilde{p_j},
\end{equation}
where the new variables satisfy the usual Poisson brackets
\begin{equation}
\{x_i,x_j\}=0,\hspace{.5cm}\{x_i,p_j\}=\delta_{ij},\hspace{.5cm}\{p_i,p_j\}=0.
\end{equation}
Using the new coordinates, we have
\begin{equation}\label{feq1}
f(r)=1-\frac{2M}{\sqrt{(x_i-\frac{\theta_{ij}p_j}{2})(x_i-\frac{\theta_{ik}p_k}{2})}}
\end{equation}
where $\theta_{ij}=1/2\varepsilon_{ijk}\theta_{k}$. The equation
\begin{equation}
f(r_H)=0,\label{horeq}
\end{equation}
where $r_H$ is the modified horizon, leads us to
\begin{equation}\label{feq2}
f(r_H)=1-\frac{2M}{\sqrt{r_{H}^{2}-\frac{\vec{L}.\vec{\theta}}{4}+\frac{P^2\theta^{2}-(\vec{P}.\vec{\theta})^{2}}{16}}}=0
\end{equation}
where $L_k=\varepsilon_{ijk} x_i p_j$, $p^2=\vec{p}.\vec{p}$ and
$\theta^{2}=\vec{\theta}.\vec{\theta}$. The Schwarzschild black
hole is non-rotating, therefore $\vec{L}=0$, in this case the new
horizon is given by
\begin{equation}\label{newcoor}
r_H=[4M^2+\frac{P^2\theta^{2}-(\vec{P}.\vec{\theta})^{2}}{16}]^{1/2}
\end{equation}
The modified Hawking-Bekenstein temperature and the horizon area
of Schwarzschild black hole in noncommutative space to second
order of $\theta$ are as following respectively
\begin{equation}\label{modtem}
T_{BH}=\frac{M}{2\pi r_{H}^{2}}=\frac{M}{2\pi
r_{0}^{2}}[1-\frac{P^2\theta^{2}-(\vec{P}.\vec{\theta})^{2}}{64M^2}]=T_0[1-\frac{P^2\theta^{2}-(\vec{P}.
\vec{\theta})^{2}}{64M^2}]
\end{equation}
\begin{equation}\label{modar}
A=4\pi r_{H}^{2}=4\pi
r_{0}^{2}[1+\frac{P^2\theta^{2}-(\vec{P}.\vec{\theta})^{2}}{64M^2}]=A_0[1+\frac{P^2\theta^{2}-
(\vec{P}.\vec{\theta})^{2}}{64M^2}]
\end{equation}
where $T_0=\frac{M}{2\pi r_{0}^{2}}$, $A_0=4\pi r_{0}^{2}$ are
Hawking-Bekenstein temperature and the horizon area in the
commutative space. The corrected entropy due to noncommutativity
of space is as
\begin{equation}\label{moden}
S=\frac{A}{4}=\frac{A_0}{4}[1+\frac{P^2\theta^{2}-(\vec{P}.\vec{\theta})^{2}}{64M^2}]=S_0
[1+\frac{P^2\theta^{2}-(\vec{P}.\vec{\theta})^{2}}{64M^2}]
\end{equation}
\section{Space noncommutativity corrections to the Cardy-Verlinde
formula} The entropy of a $(1+1)-$dimensional CFT is given by the
well-known Cardy formula \cite{Cardy} \be
S=2\pi\sqrt{\frac{c}{6}(L_0-\frac{c}{24})}, \label{car} \ee where
$L_0$ represent the product $ER$ of the energy and radius, and
the shift of $\frac{c}{24}$ is caused by the Casimir effect. After
making the appropriate identifications for $L_0$ and $c$, the
same Cardy formula is also valid for CFT in arbitrary spacetime
dimensions $d-1$ in the form \cite{Verl} \be S_{CFT}=\frac{2\pi
R}{d-2}\sqrt{E_c(2E-E_c)}, \label{cardy}
 \ee the so called Cardy-Verlinde formula, where $R$ is the radius of the system,
 $E$ is the total energy and $E_c$ is the Casimir
 energy, defined as
 \be E_c=(d-1)E-(d-2)TS.  \label{casi} \ee So far, mostly asymptotically AdS and dS
 black hole solutions have been considered \cite{AdS}-\cite{set2}. In \cite{klem},
 it is shown that even the Schwarzschild and Kerr black hole solutions, which are
 asymptotically flat, satisfy the modification of the Cardy-Verlinde formula
 \be S_{CFT}=\frac{2\pi R}{d-2}\sqrt{2EE_c}. \label{cardy1} \ee This result holds also
 for various charged black hole solution with asymptotically flat spacetime \cite{yum}\\
 In this section we compute the effect of space noncomutativity to
 the entropy of a $d=4-$dimensional Schwarzschild black hole
 described by the Cardy-Verlinde formula (\ref{cardy1}). The
 energy Eq.(\ref{qeinh}) and Casimir energy Eq.(\ref{casi}) now
 will be modified as
 \begin{equation}\label{emod}
 E=\frac{r_H}{G}=\frac{r_0}{G}[1+\frac{P^2\theta^{2}-(\vec{P}.\vec{\theta})^{2}}{64M^2}]^{1/2}
 \simeq E_0[1+\frac{P^2\theta^{2}-(\vec{P}.\vec{\theta})^{2}}{128M^2}]
 \end{equation}
 \bea\label{casmod}
 E_c&=&3E-2T_{BH}S=3E_0[1+\frac{P^2\theta^{2}-(\vec{P}.\vec{\theta})^{2}}{128M^2}]-2T_0S_0
 [1-\frac{P^2\theta^{2}-(\vec{P}.\vec{\theta})^{2}}{64M^2}]\nonumber
 \\&&
 [1+\frac{P^2\theta^{2}-(\vec{P}.\vec{\theta})^{2}}{64M^2}]
 =E_{C0}+3E_0\frac{P^2\theta^{2}-(\vec{P}.\vec{\theta})^{2}}{128M^2}.
 \eea
 We substitute expressions (\ref{emod}) and (\ref{casmod}) which
 where computed to second order in $\theta$ in the Cardy-Verlinde
 formula in order that space noncommutativity corrections to be
 considered
 \bea\label{cvmod}
 S_{CFT}=\pi r_H\sqrt{2E_c E}=\pi r_0\sqrt{2E_{c0}
 E_0}(1+\frac{P^2\theta^{2}-(\vec{P}.\vec{\theta})^{2}}{128M^2}
 [1+\frac{1}{2}(1+\frac{3E_0}{E_{C0}})])\nonumber
 \\
  =S_{0CFT}(1+\frac{P^2\theta^{2}-(\vec{P}.\vec{\theta})^{2}}{128M^2}
 [1+\frac{1}{2}(1+\frac{3E_0}{E_{C0}})]).
\eea
 \section{Conclusion}
In this paper we have examined the effects of the space
non-commutativity in the generalized Cardy-Verlinde formula. The
event horizon of the black hole undergoes corrections from the
non-commutativity of space as Eq.(\ref{newcoor}). Since the
non-commutativity parameter is so small in comparsion with the
length scales of the system, one can consider the noncommutative
effect as perturbations of the commutative counterpart\cite{li}.
Then we have obtained the corrections to the temperature and
entropy as Eqs.(\ref{modtem}, \ref{moden}). Because the
Schwarzschild black holes are non rotating, to the first order of
perturbative calculations, there is no any effect on the
properties of black hole due to the non-commutativity of space.
Then we have obtained the corrections to the entropy of a dual
conformal field theory live on flat space as Eq.(\ref{cvmod}).\\
It is necessary to mention that, our result in the present paper
is valid for a specific choice of spacetime non-commutativity
which is defined by Eqs.(\ref{alg1}, \ref{alge}). To see a more
general kind of spacetime non-commutativity refer
to\cite{{men},{chry},{ahl1}}, in these papers, the principle of
Lie algebra stability of the Poincare-Heisenberg algebra leads to
a more general kind of spacetime non-commutativity . In those
modifications, the commutators of spacetime coordinates is given
by the generators of rotations and boosts.
  
\end{document}